\documentclass{xray_symp}
\usepackage{graphics,rotating,longtable,amssymb}

\newcommand{\Ga}{$ \alpha $}
\newcommand{\Gb}{$ \beta  $}

\newcommand{\fpsfig}[1]{
\resizebox{110mm}{!}{			
\begin{turn}{270}
\includegraphics{#1}
\end{turn}
}}

\begin{document}

\title{Properties of new X-ray selected AGN
  \thanks{Visiting Astronomer, German-Spanish Astronomical Centre, Calar Alto, operated by the Max-Planck-Institute for Astronomy,
Heidelberg, jointly with the Spanish National Commission for Astronomy.}
   \thanks{based partially on observations collected at the European Southern Observatory, La Silla, Chile.}
}
\author{
K. Bischoff\inst{1} \and W. Pietsch \inst{2} \and T. Boller\inst{2} \and S. D\"obereiner\inst{2} \and W. Kollatschny\inst{1} \and H.-U. Zimmermann\inst{2}}
\mail{kbischo@gwdg.de} 
\institute{
Universit\"ats-Sternwarte, Geismarlandstra{\ss}e 11, 37083 G\"ottingen, Germany \and 
Max--Planck--Institut f\"ur extraterrestrische Physik,  Giessenbachstra{\ss}e, 85740 Garching, Germany}
\maketitle

\begin{abstract}

We present the results of a program to identify so far unknown active 
nuclei (AGN) in galaxies. Candidate galactic nuclei have been selected for optical spectroscopy from a cross-correlation of the ROSAT all sky survey (RASS) bright source catalog with optical galaxy catalogs. A high X-ray flux has been
used as pointer to galaxies with a high probability to 
contain active nuclei. Only galaxies have been accepted for the program for
which no activity was noted in NED. 
For many of the galaxies no radial velocity has been reported before. The optical spectra obtained in our first two runs demonstrate that the galaxies
cover a redshift range of 0.014 to 0.13 and that most of them
host active nuclei. For 50\% to 75\% of the candidates the X-ray emission is 
caused by the AGN.
In addition several of the remaining candidates host Seyfert 2/LINER nuclei that, however, most certainly 
are not the source of  
the X-ray emission. For a detailed analysis of the first data see Pietsch et al.\ (1998); in this paper we present results from the second optical observing run.
\end{abstract}

\section{Sample definition}

Our sample is based on a cross-correlation of the X-ray sources of the ROSAT
all-sky survey (RASS) bright source catalogue (Voges et al.\ 1996)
with the galaxies of the Catalogue of Principal Galaxies PGC
(Paturel et al.\ 1989). 
With a conservative correlation radius (maximum allowed distance
between X-ray and optical position) of 100 arcsec 
the cross-correlation resulted in a list of 1124 galaxies. The
RASS bright source catalog count rate threshold of  $\ge$ 0.05 cts/s
(corresponding to a luminosity of 
$ 2\times 10^{40}$ erg sec$^{-1}$ for a distance of 10 Mpc -- most of our
galaxies are at greater distances)
assures that all selected galaxies are clearly over-luminous in X-rays 
with respect to comparable normal galaxies (cf.\ Fabbiano et al.\ 1992).

\begin{figure}[h]
\hspace*{-5mm} \resizebox{108mm}{!}{
\begin{turn}{90} \includegraphics{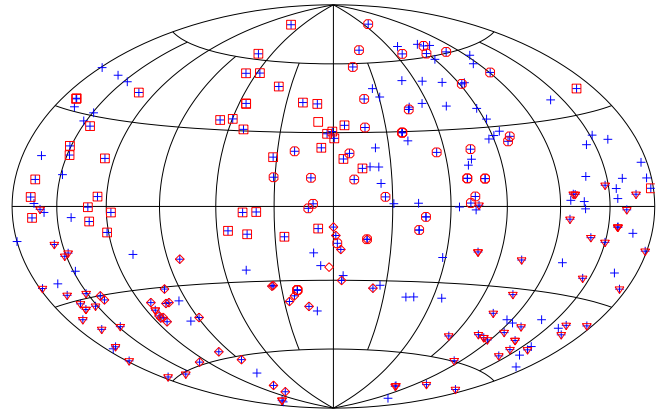} \end{turn}
}
\caption[]{Illustration of the AGN candidate sample (crosses). Open symbols mark sources already observed by optical spectroscopy in four different runs: circles are data from 1996 Nov., squares are from 1997 July, lozenges are from 1997 August, and triangles are from 1998 March.
}
\label{Fig:aitoff}
\end{figure}

To exclude known objects, for which we would expect high X-ray fluxes
(AGNs, known clusters of galaxies, nearby galaxies) 
and other candidates for the X-ray identification (stars, white dwarfs) we
checked corresponding data bases and catalogs. 
About 300 of the galaxies in our sample were 
known AGNs and for more than 250 the X-ray emission was most likely due to 
hot cluster gas. Finally we excluded objects with clearly extended X-ray 
emission which are candidates for unknown cluster emission.
Our final sample contains 230 galaxies (Fig.\ \ref{Fig:aitoff}).


\begin{table*}
\caption{Identification Information}
\begin{tabular}{lccrrll}
\hline
\noalign{\smallskip}
Name & 
RA(2000.0) & Dec(2000.0) & 
\multicolumn{1}{c}{cz(abs)} &
\multicolumn{1}{c}{cz(em)}  & AGN type & Comment \\
     & (h m s) & (d m s )                   & 
\multicolumn{1}{c}{(km s$^{-1}$)} & 
\multicolumn{1}{c}{(km s$^{-1}$)} & \\
\noalign{\smallskip}
\hline 
\noalign{\smallskip}
UGC 32              & 00 04 58.5& 11 42 03&  22260 $\pm$ 150&  22350 $\pm$ 15&Sy1.9      &                     \\
NGC 57              & 00 15 30.9& 17 19 43&   5580 $\pm$ 150&                 &non active &                     \\
NGC 70              & 00 18 22.4& 30 04 46&   7080 $\pm$ 150&   7080 $\pm$ 15&Sy2/LINER  &Arp113, group        \\
NGC 71              & 00 18 23.8& 30 03 48&   6750 $\pm$ 150&   6630 $\pm$ 15&Sy2/LINER  &Arp113, group        \\
KUG 0128+328        & 01 31 23.8& 33 08 36&                 &  21090 $\pm$ 60&Sy1        &                     \\
MCG +05-34-053      & 14 29 11.6& 30 04 38&                 &   4470 $\pm$ 60&SB         &                     \\
CGCG 163-074        & 14 32 09.0& 31 35 03&                 &  16620 $\pm$ 15&Sy1.5      &                     \\
VIIZW 608 W         & 15 32 09.7& 58 54 19&  19380 $\pm$ 150&  19200 $\pm$ 30&Sy2/LINER  &group of 3, in cl.   \\
VIIZW 608 E         & 15 32 16.2& 58 54 03&  20640 $\pm$ 150&                 &non active &group of 3, in cl.   \\
KUG 1618+410        & 16 19 51.2& 40 58 47&                 &  11370 $\pm$ 15&NLS1       &                     \\
KUG 1618+402        & 16 20 12.9& 40 09 06&                 &   8550 $\pm$ 15&Sy1        &                     \\
CGCG 196-064        & 16 26 36.4& 35 02 42&  10350 $\pm$ 150&  10260 $\pm$ 60&Sy1.5      &                     \\
NGC 6159            & 16 27 25.2& 42 40 47&   9480 $\pm$ 150&                 &Sy2/LINER  &                     \\
NGC 6160            & 16 27 41.1& 40 55 37&   9630 $\pm$ 150&                 &non active &                     \\
MCG +06-37-023      & 17 03 27.8& 36 04 19&  18840 $\pm$ 150&                 &Sy2/LINER  &                     \\
MCG +06-38-005      & 17 12 28.5& 35 53 03&                 &   8040 $\pm$ 30&Sy1.5      &asym.~Balmer profiles\\
MCG +13-12-022      & 17 18 16.6& 78 01 07&  17010 $\pm$ 150&                 &non active &cl?                  \\
MCG +05-41-010      & 17 22 15.4& 30 42 40&  13980 $\pm$ 150&                 &non active &Gpair, NE comp.      \\
NGC 6370            & 17 23 25.1& 56 58 30&   8310 $\pm$ 150&                 &Sy2/LINER  &cl?                  \\
NGC 6414            & 17 30 37.1& 74 22 32&  12600 $\pm$ 150&                 &non active &cl                   \\
MCG +07-37-018      & 18 02 40.4& 42 47 45&  15180 $\pm$ 150&                 &non active &Gpair, north.comp.   \\
                    & 18 02 39.5& 42 47 16&                 &  14070 $\pm$ 15&SB         &Gpair, south.comp.   \\
MCG +10-26-015      & 18 06 35.7& 61 35 38&   8100 $\pm$ 150&                 &non active &cl?                  \\
MCG +03-47-002      & 18 27 14.7& 19 56 19&                 &  12180 $\pm$ 15&Sy1        &                     \\
CGCG 493-002        & 21 38 33.5& 32 05 06&                 &   7380 $\pm$ 15&Sy1.5      &                     \\
CGCG 493-004        & 21 41 53.5& 31 51 28&                 &  13050 $\pm$ 30&NLS1       &cl?, very weak [OIII]\\
UGC 11950           & 22 12 31.8& 38 40 56&   6150 $\pm$ 150&                 &Sy2/LINER  &                     \\
UGC 12040           & 22 27 05.8& 36 21 42&   6390 $\pm$ 150&   6270 $\pm$ 30&Sy1.9      &                     \\
UGC 12282           & 22 58 55.4& 40 55 54&   5220 $\pm$ 150&   5130 $\pm$ 15&Sy1.9      &                     \\
UGC 12804           & 23 54 36.6& 24 33 22&  10230 $\pm$ 150&                 &non active &cl?                  \\
\noalign{\smallskip}
\hline
\noalign{\smallskip}
 & & & & & & cl = in cluster
\label{tab:identification}
\end{tabular}
\end{table*}

\section{Optical observations}

In order to cover the entire sky we performed four runs of optical observations. In November 1996 and July 1997 we used the 2.2m ESO/MPG telescope at La Silla observatory with EFOSC2 spectrograph. The data taken in November 1996 are already published (Pietsch et al.\ 1998). In August 1997 and March 1998 we carried out observations with the 2.2m telescope at Calar Alto observatory with CAFOS spectrograph. This paper deals with the data of August 1997. We used a 2048 $\times$\ 2048 SITe\#1d CCD with a pixelsize of 24$\mu m$. This setup led to a dispersion of 2 \AA\ per pixel, a spectral coverage of 4200-8200 \AA, and a spectral resolution of 8 \AA.

A publication of the data of July 1997 and March 1998 is in preparation (Bischoff et al.\ 1999).

\section{Results}

From the spectral fits we classified our objects as AGN using diagnostic 
diagrams (Fig.~2) following Baldwin et al.\ (1981).
We also determined the AGN type. 
Following the Osterbrock (1989) definition we classified the AGN in
Seyfert types according to the relative strength of narrow to broad components
of the H\Ga\ and H\Gb\ emission lines. 
%
Seyfert 2 galaxies show only narrow Balmer lines. 
Narrow line Seyfert 1 galaxies (NLS1) are a peculiar group of Seyferts
with intermediate Balmer line widths
(cf. Goodrich 1989, Osterbrock \& Pogge 1985). 
We define LINERs (Low Ionization Nuclear Emission-line Region) following Ho (1997). 
For some of our galaxies a separation between a Seyfert 2 or LINER nucleus is
not possible because [O\,{\sc iii}] and H\Gb\ 
are not detected.
Table \ref{tab:identification} lists 
the optical coordinates of our sample sources,
redshifts measured from absorption and/or emission lines, and
the determined AGN types.


\begin{figure}
\hspace*{-8mm}\resizebox{125mm}{!}{\fpsfig{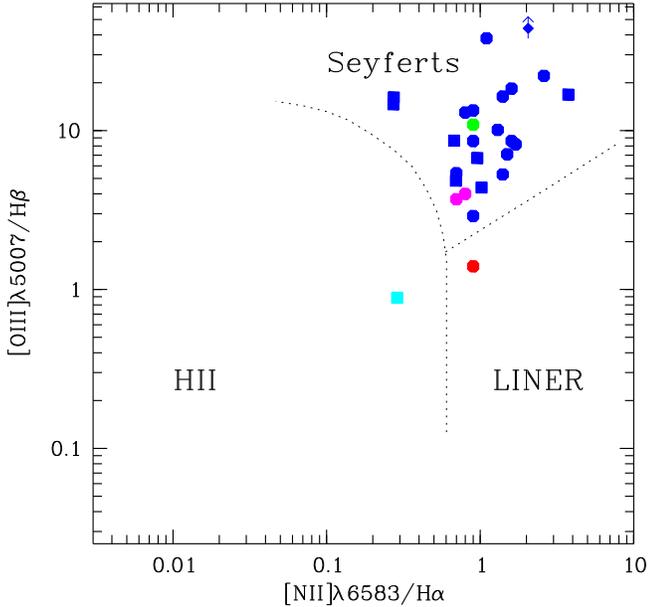}}
\caption{Diagnostic diagram of our emission-line galaxies based on a scheme of Osterbrock (1989). The dotted lines are the empirical divisions between HII galaxies, LINERs, and Seyfert galaxies. 
Circles are data from 1996 November, squares are from 1997 August.
The arrow indicates an upper limit of H\Gb.
}
\label{fig:baldwin}
\end{figure}

\begin{table}
\caption{AGN type distribution. The described classification scheme resulted in the following distribution of sources. 
}
\label{AGNdistr}
\begin{tabular}{lccccc}
\hline
\noalign{\smallskip}
                & 96 Nov. & 97 Aug. & total \\
                &        &        & frequency\\
\noalign{\smallskip} \hline \noalign{\smallskip}
Sy1             & 5      & 3      & 11\% \\
Sy1.2 \dots 1.9 &16      & 7      & 33\% \\  
NLS1            & 1      & 2	  & ~4\% \\
Sy2             & 2      &--	  & ~3\% \\
LINER           & 3      &--	  & ~4\% \\
Sy2/LINER       & 4      & 7	  & 16\% \\
BL Lac          & 3      &--	  & ~4\% \\
QSO             & 1      &--	  & ~1\% \\
H{\sc ii}       & --     & 2	  & ~3\% \\
non active      & 3      & 9	  & 16\% \\
\noalign{\smallskip}
\hline
\end{tabular}
\end{table}

To analyse the radio properties of our sources we looked for counterparts in recent radio surveys. The 1.4 GHz radio flux was taken from the NRAO VLA Sky Survey NVSS (Condon et al.\ 1998), the 4.85 GHz flux from the Parkes-MIT-NRAO PMN survey (Griffith et al.\ 1995).

Figure \ref{fig:ratios} (left panel) shows the $f_r / f_B$ ratio for objects with radio counterparts. 
Using the formalism of Kellermann et al. (1989) 
one can quantify the radio-quiet or radio-loud nature of our objects.
The ratio of radio to optical flux density for most objects is below 10 
and objects below such a value are considered as radio-quiet (of course, there is no strict dividing line between radio-quiet and radio-loud objects). 
The most intense radio emitters in our sample (RX~J011232.8-320140 and NGC~1218) show a radio to optical flux ratio of about 275.

From the RASS count rate we calculated the X-ray flux assuming a power-law spectrum with fixed photon spectral index $\Gamma=2.3$,
which is the typical value found for extragalactic objects with ROSAT
(cf. Hasinger et al. 1991, Walter \& Fink 1993), and
an absorbing column density fixed at the individual
Galactic hydrogen value N$_{\rm H gal}$ along the 
line of sight (Dickey \& Lockman 1990). 
With this procedure we do not correct for absorption of the X-rays within the galaxy. Therefore, our X-ray fluxes derived with this method have to be understood as lower limits of the intrinsic fluxes if the X-rays are emitted from active nuclei within the galaxies. 

The optical B magnitudes were taken from the NASA Extragalactic Database (NED)\footnote{The NASA/IPAC Extragalactic Database is operated by the Jet Propulsion Laboratory, California Institute of Technology, under contract with the National Aeronautics and Space Administration.}.
If no B magnitude was available in NED we took the magnitude from the PGC (Paturel et al.\ 1989). For VII\,Zw\,608\,E (eastern component) and the companion of MCG+07-37-018 the magnitudes were estimated from our acquisition images.
The monochromatic flux at 4400\AA\ $f_B$ and the total far-infrared (40-120$\mu$m) fluxes $f_{\rm FIR}$ were calculated as described in Pietsch et al.\ (1998). To compute absolute magnitudes and luminosities we adopted a Hubble constant 
$H_0 = 75~ \rm km\ s^{-1}\ Mpc^{-1}$ and cosmological deceleration parameter 
of $\rm q_0 = 0.5$.

\section{Discussion}

Most of our sources turn out to harvest an active nucleus or even
merging nuclei where at least one component is active. 
It is, however, not yet clear if these active nuclei 
really are the source of the X-rays.

We have several ways to attack this problem and can propose
solutions for individual galaxies:
\begin{itemize}
\item Information about source extension from the RASS catalog: If a source is extended the major part of the X-ray emission will not originate in the nucleus (8\% of the sources). 

\item Unresolved HRI detection of a source centered on the nucleus
strongly argues for a nuclear origin of the X-rays (24\% of the sources).

\item Time variability in the X-ray flux rules out emission from
extended gas clouds and strongly argues for an AGN (24\% of the sources).

\item 
Comparision of measured X-ray luminosity with those expected from the AGN / morphology type of the galaxy or for the galaxy surroundings (group or cluster 
environment) can give hints on the origin of the X-rays.
In some cases this would argue for an origin of the X-ray emission 
in the surrounding group or cluster (16\% of the sources); 
Hot gas in the early type host galaxy could explain the X-ray emission for others (13\% of the sources).

\item 
Line fitting analysis of prominent optical emission lines: The location of our objects in the diagnostic diagrams of Osterbrock (1989) clearly demonstrates their AGN nature (cf.~Fig.\ \ref{fig:baldwin}) (47\% of the sources).
\end{itemize}

\begin{figure*}
\hbox{
\hspace*{-6mm}
\resizebox{82mm}{!}{\fpsfig{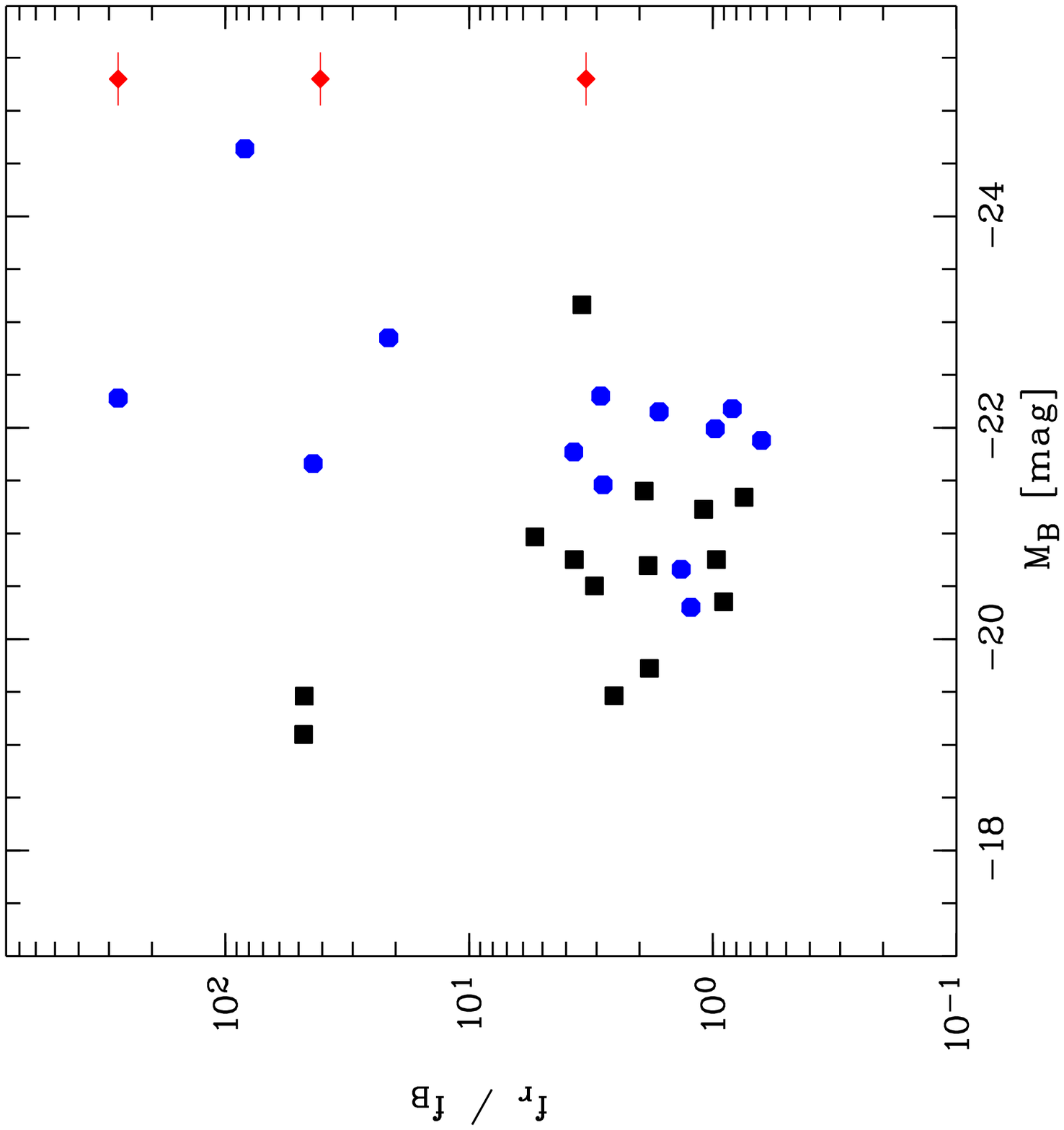}}
\hspace*{-24mm}
\resizebox{82mm}{!}{\fpsfig{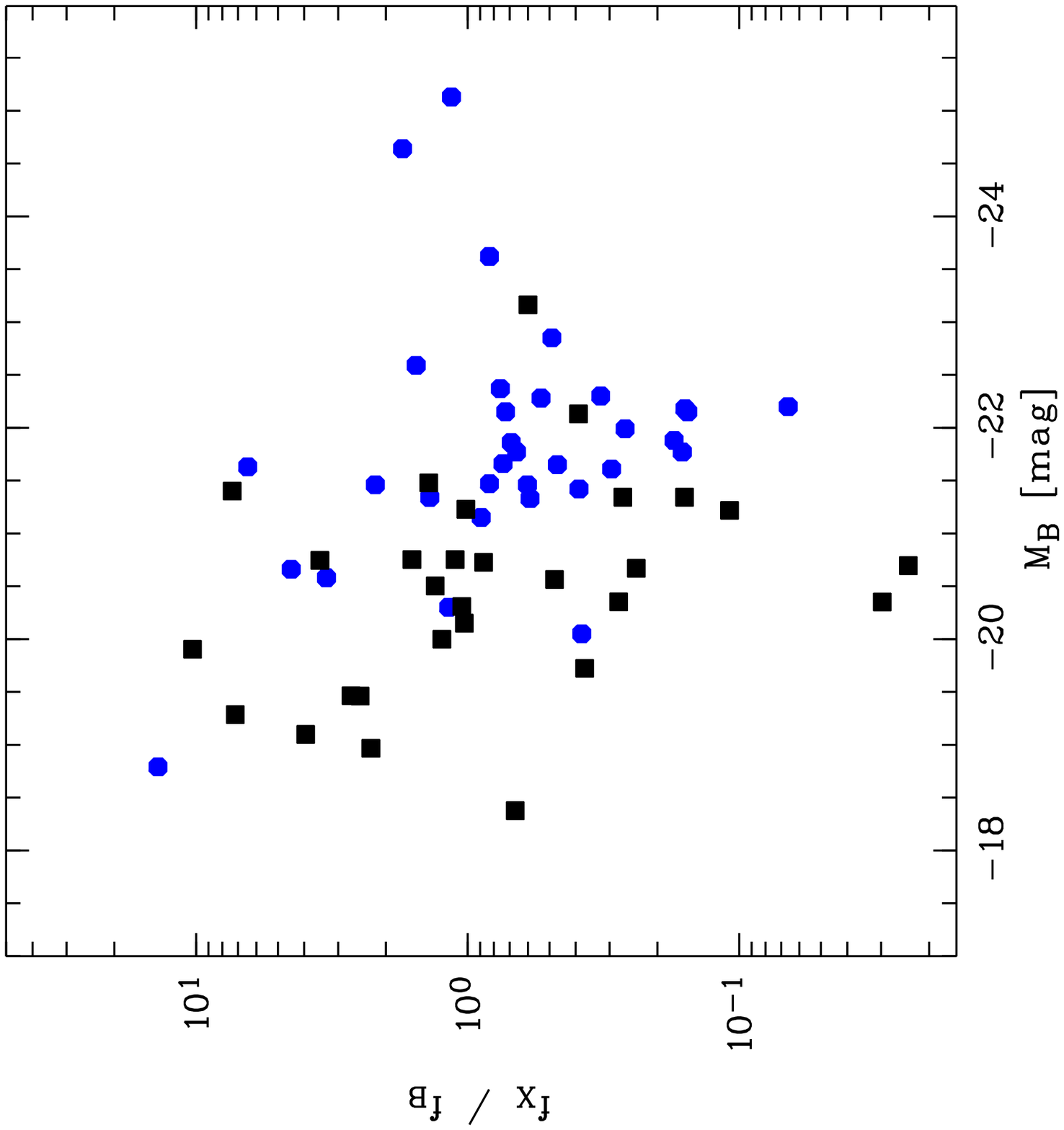}}
\hspace*{-24mm}
\resizebox{82mm}{!}{\fpsfig{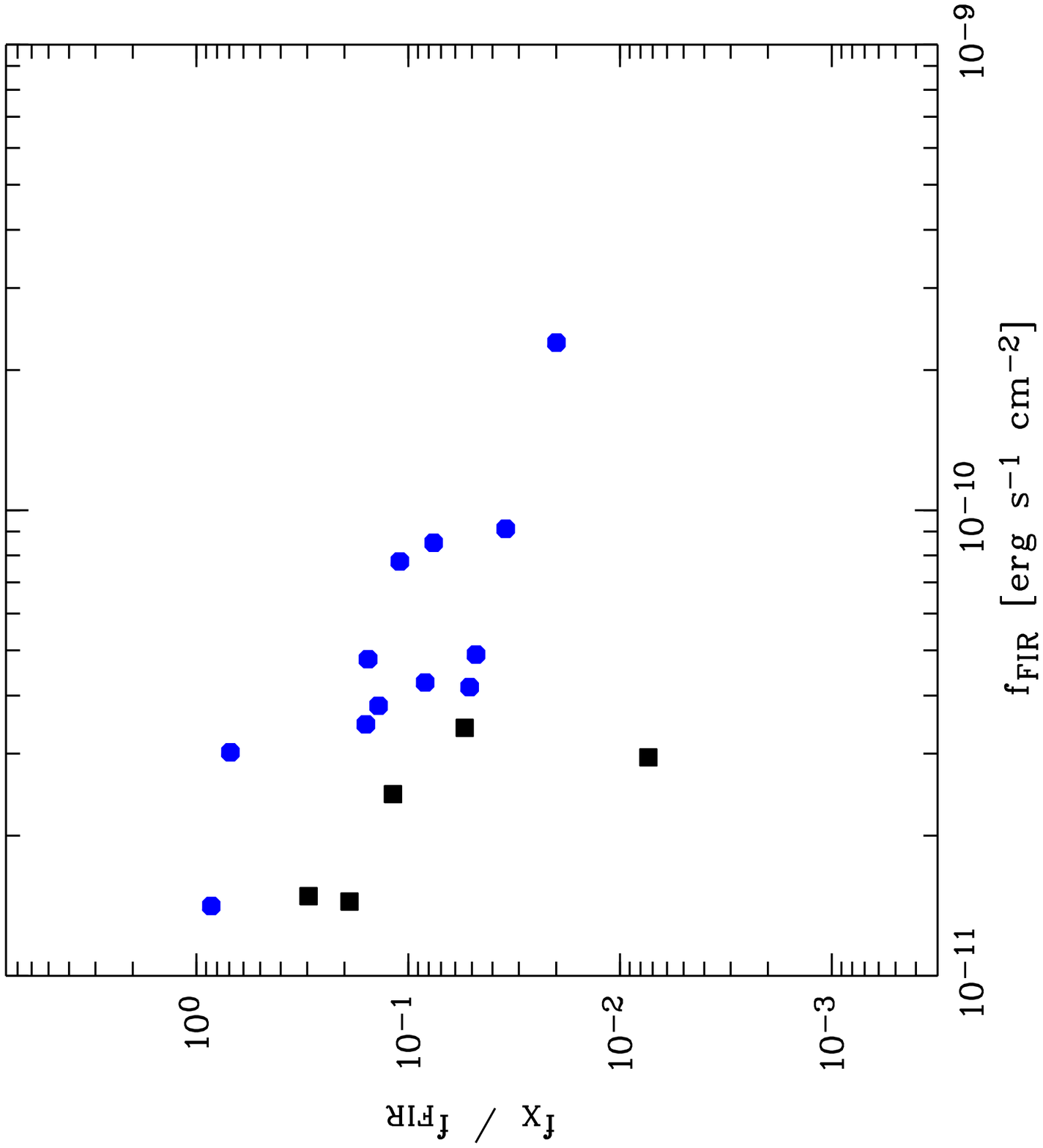}}
}
\caption{
Flux ratios as functions of absolute B magnitude M$_B$ or far infrared flux f$_{FIR}$, respectively.
Left: The f$_r$/f$_B$ ratio is used to discriminate radio-loud and radio-quiet objects. The 3 
lozenges plotted at M$_B$ = -25.3 are the BL Lac objects with unknown absolute optical magnitude M$_B$.
Middle: Comparing with
the nomograph shown in Fig.~1 of Maccacaro et al.\ (1988),
one sees that the combined X-ray flux and optical M$_{\rm \bf B}$ 
agree well with those expected from AGN. 
Right: 
All objects besides one show flux ratios f$_X$/f$_{FIR}$ above a
value of 0.01. 
Theoretical models 
suggest that flux ratios above about 0.01 require
AGN activity, whereas values below about 0.01 can be explained by
starburst activity. 
Circles are data from 1996 November, squares are from 1997 August.
}
\label{fig:ratios}
\end{figure*}

The flux ratios are an additional indication of the AGN character of the objects. Using the nomograph shown in Fig.~1 of Maccacaro et al.\ (1988),
one sees that the combined X-ray flux and optical M$_{\rm B}$ 
of our objects agree well with those expected from AGN (Fig.\ \ref{fig:ratios}, middle panel).
The right panel gives the ratio of the X-ray to far-infrared
(40-120 $\mu m$) flux. All of the objects show flux ratios above a
value of 0.01. As shown in Fig.~5 of Boller et al.\ (1998) most AGN
exhibit flux ratios above that value. Theoretical models addressing the
X-ray and far-infrared emission of galaxies in different states of
nuclear activity suggest that flux ratios above about 0.01 require
AGN activity, whereas values below about 0.01 can be explained by
starburst activity (Bertoldi \& Boller, in preparation).

The results of our optical follow up observing runs have demonstrated
that our selection strategy from the RASS bright source catalog -- PGC
correlations is rather efficient in detecting new active galactic nuclei. 
From all observations we expect the detection of about 100 new members of the nearby AGN population. The detailed analysis of the entire data and the discussion of its implications will be published in a following paper.

\begin{acknowledgements}
The ROSAT project is supported by the German Bundesministerium f\"ur
Bildung, Wissenschaft, Forschung
und Technologie (BMBF/DLR) and the Max-Planck-Gesellschaft (MPG).
This work has been partially supported by 
Deutsches Zentrum f\"ur Luft- und Raumfahrt e.V.\ (DLR)
grant 50\,OR\,9408\,9.
\end{acknowledgements}

\end{document}